\documentclass[12pt]{article}
\usepackage[english]{babel}
\usepackage{a4wide}
\date{}
\begin{document}

\title{\bf {\large{Spinors and Octonions}}}
\author{\normalsize{Luis J. Boya \footnote{To Alberto Galindo in his
seventieth birthday}}  \\
\small{luisjo@unizar.es}
\\ \\ \\
\normalsize{Departamento de F\'{\i}sica Te\'orica} \\
\normalsize{Universidad de Zaragoza} \\
\normalsize{E-50009 Zaragoza} \\
\normalsize{SPAIN}}

\maketitle

\begin{abstract}

Octonions are introduced through some spin representations. The
groups $G_2$, $F_4$ and the $E$ series appear in a natural manner;
one way to understand octonions is as the ``second coming" of the
reals, but with the spinors instead of vectors. Some physical
applications in $M$- and $F$-theory as putative ``theories of
everything" are suggested.

\end{abstract}

\section{The Seven Sphere}.

Consider the real, complex and quaternion numbers $R$, $C$, $H$.
Identify the normed vector spaces  $R^{8k} \cong C^{4k} \cong
H^{2k}$, and write the natural inclusion of the isometry groups

\begin{equation} \label{eq:1}
O(8k) \supset U(4k) \supset Sp(2k)
\end{equation}

Recall now the \textit{Spin groups}, which cover the rotation
groups twice, $Spin(n)/Z_2=SO(n)$; remind only that $Spin(7)$ has
a real 8-dim irreducible representation; as $SO(n)$ is the maximal
isometry group for spheres, the previous sequence becomes for
$k=1$

\begin{equation} \label{eq:2}
 \begin{array}{ccccccc}
 Sp(2) &\subset & SU(4) &\subset & Spin(7) & \subset & SO(8) \\
H & \ & C & \ & (O) & \ & R
\end{array}
\end{equation}

\noindent where the adscription of division algebras other than
$(O)$ is clear. It is fairly easy to see that all these group act
\textit{trans} on the seven sphere of constant norm vectors in $R^
8$; therefore after finding the stabilizers we get

\begin{equation} \label{eq:3}
SO(8)/SO(7) = Spin(7)/G_2=SU(4)/SU(3) =Sp(2)/Sp(1) = S^7
\end{equation}

\noindent where $G_2$ is \textit{defined} to be the little group
of $Spin(7)$ acting on $S^7$; the other isotropies are obvious.
From the above inclusions (\ref{eq:2}) we get \textit{four}
conmutative
diagrams \\

O vs. R

\begin{equation} \label{eq:4}
 \begin{array}{ccccc}
 G_2 &\rightarrow & Spin(7) &\rightarrow & S^7  \\
 \downarrow &  & \downarrow &  & \vert \vert  \\
SO(7) & \rightarrow & SO(8) & \rightarrow & S^7 \\
\downarrow &  & \downarrow & &  \\
RP^7 & = & RP^7 &  &
\end{array}
\end{equation}

\noindent where we learn from the first column an interesting
result. \\

H vs. C

\begin{equation} \label{eq:5}
\begin{array}{ccccc}
 Sp(1) &\rightarrow & Sp(2) &\rightarrow & S^7  \\
 \downarrow &  & \downarrow &  & \vert \vert  \\
SU(3) & \rightarrow & SU(4) & \rightarrow & S^7 \\
\downarrow &  & \downarrow & &  \\
S^5 & = & S^5 &  &
\end{array}
\end{equation}

\noindent where the vertical lines are obvious: $SU(3)/SU(2) = S^5
= Spin(6)/Spin(5)$: we learn \textit{passim} the Cartan identies

\begin{equation} \label{eq:6}
Sp(1)=Spin(3)=SU(2), \ \ \     Spin(5)=Sp(2) \ \ \    and \ \ \
Spin(6)=SU(4)
\end{equation}

 C vs. O

\begin{equation} \label{eq:7}
 \begin{array}{ccccc}
 SU(3) &\rightarrow & SU(4) &\rightarrow & S^7  \\
 \downarrow &  & \downarrow &  & \vert \vert  \\
G_2 & \rightarrow & Spin(7) & \rightarrow & S^7 \\
\downarrow &  & \downarrow & &  \\
S^6 & = & S^6 &  &
\end{array}
\end{equation}

\noindent and we learn of the 7-dim irrep of $G_2$, as $S^6
\subset R^7$. From (\ref{eq:3}) we already know dim $G_2 =14$. Now
the dimension of 3-forms in $R^7$ are ${7 \choose 3}= 35 = 7^2
-14$ : hence $G_2$ leaves invariant a generic three-form in $R^7$
(this becomes the octonion multiplication, of course, and with a
little extra effort we conclude that $G_2$ is the automorphism
group of a multiplicative structure: because of the metric, we
trade the 3-form $T_3^0$ by a $T_2^1$ tensor, which characterizes
algebras;
this is our way of introducing octonions!). \\

There is a fourth diagram \\

C vs. R

\begin{equation} \label{eq:8}
 \begin{array}{ccccc}
 SU(3) &\rightarrow & SU(4) &\rightarrow & S^7  \\
 \downarrow &  & \downarrow &  & \vert \vert  \\
SO(7) & \rightarrow & SO(8) & \rightarrow & S^7 \\
\downarrow &  & \downarrow & &  \\
M_{13} & = & M_{13} &  &
\end{array}
\end{equation}

 \noindent where  $M_{13}$ is a symmetric space. We shall
see that there is sense in calling $Spin(7) \sim Oct(1)$. \\

 The relation (\ref{eq:1})  holds for any $k$. For $k=2$ it is expanded to

 \begin{equation} \label{eq:9}
 \begin{array}{ccccccc}
 Sp(4) &\subset & SU(8) &\subset & Spin(9) & \subset & SO(16) \\
 H &  & C & & (O) & & R
\end{array}
\end{equation}

\noindent as, again, dim $Spin(9)=16$. We shall only reproduce the
$Spin(9) = ``Oct(2)" \subset O(16)$ relation for $k=2$ \\

O vs. R

\begin{equation} \label{eq:10}
 \begin{array}{ccccc}
Spin(7) &\rightarrow & Spin(9) &\rightarrow & S^{15}  \\
 \downarrow &  & \downarrow &  & \vert \vert  \\
SO(15) & \rightarrow & SO(16) & \rightarrow & S^{15} \\
\downarrow &  & \downarrow & &  \\
M_{84} & = & M_{84} &  &
\end{array}
\end{equation}

We do not know of any interpretation of the $M_{84}$ manifold,
except that reminds one of the 3-forms in 11 dimensions (9
effective) associated to the membranes in $M$ theory. We shall
also
see that there is a sense in calling $ Spin(9) \sim Oct(2)$. \\

For a good reference on Spinor groups look at Porteous\'{}s book
\cite{Port}. \\

\section{Octonions}
Granted that the multiplicative structure we found before is
invertible, but nonassociative, we have the division
algebra of octonions $(O)$. We know also \\

i)  The automorphism group of $(O)$ is $G_2$. There is a natural
7-dim representation acting upon the imaginary octonions (of
course, the real part is elementwise invariant); also, $G_2$
respects the octonion norm, hence the 6-sphere remains invariant:
this fact leads up to endowing the 6-sphere with a
(quasi!)-complex structure (no complex, because no symplectic: the
spheres are 2-connected!). The construction of quaternions can be
carried out exactly equal: in $R^3$ live the imaginary
quaternions, and imaginary quaternion product (vector product)
becomes a 3-form, hence a volume form. The invariance group is
$SL(3, R)$, but there is also an invariant norm, hence

\begin{equation} \label{eq:11}
Aut \ H = SL(3, R) \  \cap \  O(3) = SO(3)
\end{equation}

ii) If the octonion division algebra were associative, $S^7$ will
be $Oct(1)$, in the sense of the previous true relations

\begin{equation} \label{eq:12}
O(1) = S^0, \ \ \    U(1) = S^1, \ \ \     Sp(1) = S^3,
\end{equation}

But here $\grave{}Oct(1)\acute{}$ = $S^7$ gets stabilized through
the automorphism group $G_2$ to become the \textit{twisted} sphere
structure $Spin(7) \cong  S^3(\times S^7 (\times S^{11}$ or

\begin{equation} \label{eq:13}
Oct(1) = Spin(7)
\end{equation}

I explained these twists in \cite{Boy1}. Let us look for the
projective line $OP^1$. For the other algebras we have

\begin{eqnarray*}
RP^1 = S^1= O(2)/O(1)^2, \ \   CP^1= S^2=U(2)/U(1)^2,
\end{eqnarray*}
\begin{equation} \label{eq:14}
\ \   HP^1= S^4=Sp(2)/Sp(1)^2
\end{equation}

But here we get, instead

\begin{equation} \label{eq:15}
OP^1 = S^8 = Spin(9)/Spin(8) = Oct(2)/Oct(1)^2
\end{equation}

Does it make sense to call $Oct(1)^2 \sim Spin(8)$? Yes:
$Oct(1)^2$ should be $S^7 \times S^7$ stabilized by $G_2$ again;
and indeed in twisted spheres

\begin{equation} \label{eq:16}
Spin(8) = S^3 (\times S^7 (\times S^7 (\times S^{11}, \ \  \ \
Spin(9) = S^3 (\times S^7 (\times S^{11} (\times S^{15}
\end{equation}

\noindent as sphere structure, so also $Spin(9) \sim Oct(2)$. \\

iii) To complete the issue, we reach $Oct(3)$ but not more,
because nonassociativity prohibits this (the mathematical reason
is told below). It turn out that

\begin{equation} \label{eq:17}
Oct (3) = F^4  \cong S^3 (\times S^{11}(\times S^{15} (\times
S^{23}
\end{equation}

\noindent  and

\begin{equation} \label{eq:18}
OP^2 = F_4 /B_4  \cong  S{\cdot}Oct(3)/Oct(2) = F_4/Spin(9)
\end{equation}

So the factor of $\grave{}(Oct 1)\acute{}$ dissapears as $O(1)$
disssapears in

\begin{equation} \label{eq:19}
RP^n = O(n+1)/O(n)\times O(1) = SO(n+1)/O(n)
\end{equation}

All this reinforces the octonions as : 1) The second coming of the
reals, after dimension 8; and also 2) With spinors, not with the
vectors of the orthogonal group. \\

For general introduction of octonions for physicist see \cite{Ram}
and \cite{Gur}; some of the identifications above are in
\cite{Boy2}. \\

\section{The Magic Square}
 The natural inclusions, for any $n$

\begin{equation} \label{eq:20}
 \begin{array}{ccccc}
O(n) &\subset & U(n) &\subset & Sp(n)  \\
  &  &  & &  \cap \\
  &  &  & &  U(2n) \\
  &  &  & &  \cap \\
    &  &  & & O(4n)
\end{array}
\end{equation}

\noindent are the key to understand symmetric spaces, as we have
shown elsewhere \cite{Boy2}. There are seven classes of
(\textit{classical}) symmetric spaces, the four associated to the
previous diagram, namely

\begin{equation} \label{eq:21}
U(n)/O(n),  \ \ Sp(n)/U(n), \ \  U(2n)/Sp(n) \ \  and \ \
O(2n)/U(n)
\end{equation}

\noindent and the three families associated to different ``floors"
$n=p+q$:

\begin{equation} \label{eq:22}
K(p+q)/K(p)\times K(q), \ \     with \ \  K= R, C \  or \ H,
\end{equation}

\noindent which are all grassmannians (projective spaces for
$q=1$). Cartan found all this around 1926, and he even went
further and classified the exceptional symmetric spaces; the
standard source is \cite{Hel}. \\

To understand these, recall the octonions, although
nonassociative, are alternative: the associator of three octonions

\begin{equation} \label{eq:23}
[a, b, c] := a(bc) -(ab)c
\end{equation}

\noindent is fully antisymmetric. Therefore the symmetric algebra
is a (conmutative) Jordan algebra. \textit{The exceptional groups
(except $G_2$) are automorphism groups of certain Jordan algebras
over the octonions.} \\

We form first the mutilated square for the first floor in
(\ref{eq:20})

\begin{equation} \label{eq:24}
 \begin{array}{ccccccccc}
 O(1) &\subset & U(1) &\subset & Sp(1) & \subset & Oct(1) & = & Spin(7) \\
 \cap &  & \cap & & \cap & & & & \\
 U(1) & \subset & U(1)^2 & \subset & U(2) & & & \\
 \cap &  & \cap & & \cap & & & & \\
 Sp(1) & \subset & U(2) & \subset & O(4) & & &
\end{array}
\end{equation}

\noindent extending it in a natural way. For $n=2$, it is
completed:

\begin{equation} \label{eq:25}
 \begin{array}{ccccccccc}
 O(2) &\subset & U(2) &\subset & Sp(2) & \subset & Oct(2) & = & Spin(9) \\
 \cap &  & \cap & & \cap & & & & \cap \\
 U(2) & \subset & U(2)^2 & \subset & U(4) & \subset  & & & Spin(10) \\
 \cap &  & \cap & & \cap & & & & \cap \\
 Sp(2) & \subset & U(4) & \subset & O(8) & \subset & & & Spin(12) \\
 \cap &  & \cap & & \cap & & & & \cap \\
Spin(9) & \subset & Spin(10) & \subset & Spin(12) & \subset & & &
Spin(16)
\end{array}
\end{equation}

The last column can be understood as complexification,
quaternionization and octonionization of  $Spin(9) = Oct(2)$; see
Freudenthal \cite{Freu} or \cite{Gur}. \\

And now the exceptional groups are obtained from the last
column/row: as we generate $O(n+1)$ by adding the vector irrep to
the adjoint irrep (check ${n+1 \choose 2} = {n \choose 2} + n)$,
we trade
vector by spinor: \\

$F_4$ is the extension of $Spin(9)=Oct(2)$: adjoint + spin,
$36+16=52$ \\

As we get $SU(n+1)$ from $SU(n)$ by adding the vector and a $U(1)$
factor to the adjoint $((n+1)^2 -1 = n^2 -1 + 2n+1)$, here we get
$E_6$: \\

$E_6$ is the extension of $O(10)$ with the $Spin + U(1):
45+1+32=78$ \\

As we get $Sp(n+1)$ from $Sp(n)$ by adding the vector and a
$Sp(1)$ factor, here we get $E_7$: \\

$E_7 $ extends $O(12): adj + Sp(1) + spin$,  \ \      $66+3+64=
133$ \\

Finally, the octonionic extension requires only the spin irrep, as
octonions behave like the reals: \\

$E_ 8 $ \ \ extends  \ \            $O(16): adj + spin;   \ \ 120
+
128 = 248$ \\

One checks the skew square of the spin irrep contains the adjoint,
and that the Jacobi identity is satisfied ( a nontrivial task!).
\\

This allows us to write the \textit{Freudenthal magic square} in
the conventional form

\begin{equation} \label{eq:26}
 \begin{array}{cccc}
 O(3)  & U(3)   & Sp(3) &  F_4 \\
U(3) &  U(3)^2 &  U(6)  &   E_6 \\
Sp(3) &  U(6)  &  O(16)  &  E_7 \\
F_4    &  E_6     &    E_7    &    E_8
\end{array}
\end{equation}

So we see that the exceptional groups (except $G_2$) are
extensions of some orthogonal groups by the spin representation:
the character $O$, $U$, $Sp$ or $Oct$ is reflected in the $0$,
$U(1)$, $Sp(1)$, $0$ added factors. For $G_2$, it is the little
group of $Spin(7)$ acting in the seven sphere, as said. This
relation between exceptional groups and spinors, which I learned
from \cite{Adams}, remains a bit mysterious. \\

The symmetric spaces involving the exceptional groups are now very
clear; there are twelve of them. We shall only exhibit the four
associated with $E_6$. They are the quotients in the graphs
(\ref{eq:25}) and (\ref{eq:26})

\begin{equation} \label{eq:27}
 \begin{array}{ccccc}
&  & F_4   &  &  Spin(10) \\
 &   &  \cap  &  & \cap \\
SU(6) &  \subset & E_6  &  =  &  E_6 \\
\end{array}
\end{equation}
\noindent and account for three of them, the fourth is associated
to the split form, and it is $E_6/Sp(4$). We shall consider one of
them:

\begin{equation} \label{eq:28}
E_6/O(10){\cdot}U(1)
\end{equation}

\noindent that is, the complexification of the Moufang plane, with
$32$ dimensions. It is the simplest (rank two) hermitian
exceptional
symmetric space. \\

\section{Supersymmetry and F-Theory}

 After this excursion in
pure dry mathematics, it is healthy to inject some physics. The
idea is
that projective geometry could play a role in the real world! \\

    The Moufang octonionic plane considered before

\begin{equation} \label{eq:29}
OP^2 = F_4/B_4 = F_4/Spin(9)
\end{equation}

\noindent ends the series $RP^2 = SO(3)/O(2), CP^2 = SU(3)/U(2)$,
and $HP^2 = Sp(3)/Sp(1){\cdot}Sp(2)$.  Projective spaces of dimension
higher than two are necessarily desargian, which implies the
underlying number field is associative; hence, the octonion groups
stop at three. \\
Now for some physics, in concrete 11-dim supergravity. P. Ramond
\cite{Ram2} has shown that the maximal supergravity multiplet
(triplet) in 11 dimensions (9 transverse)

\begin{equation} \label{eq:30}
 \begin{array}{ccc} graviton \  h & - gravitino \ \psi & + 3-form \ C \\
 44 & - 128 &  + 84
\end{array}
\end{equation}

\noindent corresponds to the three embedings of $B_4$ in $F_4$ :
this ``3" is precisely the Euler number of $OP^2$ (with homology
only in zero, eight and 16 dimensions, $b_0 = b_8 = b_{16} = 1$,
others =0)); senior physicists will recall the three embedings of
SU(2) in flavour SU(3), as Isospin, U-spin and V-spin. As eleven
dimensional gravity is supposed to be the low energy limit of
M-theory, we see that in this incompletely known theory octonions
and a projective plane already play a role. \\

The mathematician Kostant \cite{Kost} has proven that this is
general phenomenon in coset spaces X = G/H where G is a semisimple
Lie group and H reductive, with $G$ and $H$ of the same rank.
Namely the identity representation of G induces $\chi$ irreducible
representations of H, where $\chi$ is the Euler number of the
coset space X; this number is also the quotient of the orders of
the Weyl reflection groups of G resp. H. These representations
arrange themselves as the Spin representation of the SO(dim X )
group; here dim X = 52 - 36 =16, and  the result of Kostant is, in
our case

\begin{eqnarray*}
Spin_L(16) - Spin_R(16) = h +  C  -  \psi
\end{eqnarray*}
\begin{equation} \label{eq:31}
128    -     128       = 44 + 84 - 128
\end{equation}

Many supermultiplets (not all) can be understood in this way, see
also \cite{Brink}. This links up supersymmetry, spinors and
octonions in an intrincate way, not well elucidated up to now. \\

As is it well known, the 11D Sugra multiplet is too small to
encompass the spectrum of the standard model. So the question
arises whether there is a different coset space furnishing a more
realistic supersymmetric multiplet. \\

    We claim the former space (\ref{eq:28}) or more precisely

\begin{equation} \label{eq:32}
Y: = \frac{E_6/Z_3}{Spin^c(10)}
\end{equation}

\noindent (where $Z_3$ is the center of $E_6$ and $Spin^c(n):
=Spin(n) \times _{Z_{2}} U(1)$) is a better candidate, although it
contains, on the face of it, too many states. The plane Y
corresponds precisely to the complexification of the Moufang plane
\cite{Atiyah}, and the Euler number is

\begin{equation} \label{eq:33}
\chi (Y) = \# Weyl \ group \ of E_6 / \# Weyl \ group \  of D_5 =
 51840/1920    = 27
\end{equation}

Therefore the Id irrep of $E_6$ generates, in a supersymmetric
fashion,
 27 $2^{16}$-dim irreps of $SO(10) \times U(1)$!  They are obtained by the
 skew products of the Spin irrep of SO(10),
 of complex dimension 16 (with an irrelevant "charge" label associated to the U(1) subgroup).
 The total splitting had been calculated by I.
 Bars \cite{Bars} in another context, and we just write it for the record:

\begin{equation} \label{eq:34}
(1 - 1)^{16} = 1 - 16 + {16 \choose  2} - {16 \choose 3}. . . \pm
. . . - {16 \choose 15} + 1 = 32 768 -  32 768
\end{equation}

\noindent corresponding to the p-forms in $C^{16}$ (the Spin(10)
irrep is complex). The split with respect to $O(10) \times O(2)$
is

\begin{equation} \label{eq:35}
\begin{tabular}{lllll}
\hline\noalign{\smallskip}
SU(16) & Spin(10) & O(9) & O(8) & F-Th \ particle \\
\noalign{\smallskip}\hline\noalign{\smallskip}
 1 & 1 & 1 & 1 & scalar \\
 -16 & 16 & 16 & $8_L + 8_R$ & spinor \\
+120  & 120  & 84+36 & 56+28, 28 + 8 + ...  & 3-form \\
  -560 & 560    & 432 + 128    &   ...     &  hypergravitino \\
   1820  & 770+ 1050  &   924 + ...   & ... & Weyl Tensor \\
   & & & & + self-6-form  \\
   -4368     &  3696 + 672  &  2560+ ... +672  &  & \\
   8008  &     4312+3696   &  2457+... & & \\
   -11440  &     8800+2640  &  5040+... & & \\
   +12870   &    4125+8085  &  3900 +... & & \\
   &  + 660 & &  & \\
\noalign{\smallskip}
\end{tabular}
\end{equation}

\noindent plus the conjugate irreps for the $k=9$ to 16 skew
tensors
of SU(16). \\

There is no simple way to relate these representations to known
particles; there must be a particular truncation, different from
the naive square root (which will reproduce conventional 11D
Sugra) from $2^{16}$ to $256 = 2^8$ states, as required by the
minimal supersymmetric model. We only remark here that among the
spectrum of  $O(10) \times U(1)$ multiplet neither the graviton
(55-dim) nor the gravitino (144-dim) appear, in agreement (if
vaguely) with the idea that the supersymmetry is realized first
without gravitation; in this sense $F$-theory (in 12 dimenions)
fares better than $M$ (in 11). There is also the feature than
$E_6$ plays a role, and it is the maximal group well fitted for
Grand Unified Theories, GUTs. And, for numerologists, dim Poincar\'{e}
$\sim$ de Sitter
/AdS in (2, 10) space = dim $E_6$ = 78. \\

Finaly, there is another intriguing feature, related to the
mentioned square root: The Sugra 11D multiplet can be seen as the
"square" of the Yang-Mills multiplet in ten dimensions (8
effective) \\

 $(vector - spin_L)\times (vector - spin_R) = 44 + 1 + 28 + 8 + 56 - 8 -8 -56
 -56$ \\

\noindent that is, the graviton+dilaton+vector+2-form+3-form Bose
content of IIa Sugra in 10D +plus fermions (two spinors (8) plus
two gravitinos (56)). Now this "oxidises" to 11d N=1 Sugra as
known: \\

graviton + dilaton + vector \  in \  10D = graviton \  in \  11D \
(44=35+8+1)
\\

$2- \ \&  \ 3-form \  in \  10D = 3-form \  in \  11D  \ (84=56+28)$ \\

$L \ \& \ R \ spinor + gravitinos \  in \  10D = graviton \  in \
11D \ (2{\cdot}8 +2.56=128)$
\\

Now Bars \cite{Bars} has shown that another square produces the
supermultiplet in (2, 10) dimensions!  \\

In toto \\

$[(vect - spin_L)\times (vect - spin_R)]^2 = Y-multiplets \ of \
O(10) \times O(2)$ \\

$[(8 - 8) \times (8 -8)]^2  =    (16 \times 16)^2 = 32768
- 32768$ \\

So it seems that the natural extension of the fundamental
supersymmetric 10D multiplet, given by the triality of O(8), gives
in the fourth power the supermultiplet of the symmetric space Y in
12D, but now with two times; this also shows, \textit{inter alia},
that IIB
string theory fits naturally in F-theory. \\

 We leave this at that; we have not succeeded in truncating the enormous multiplet
 in a realistic way; but the issue is worth pursuing...

\end{document}